\documentstyle[aps,twocolumn,prl,epsf]{revtex}

\begin{document}

\title{Effects of dilute Zn impurities on the uniform magnetic
susceptibility of YBa$_2$Cu$_3$O$_{7-\delta}$ \\}

\author{ N.\ Bulut }

\address{
Department of Mathematics, Ko\c{c} University, 
Istinye, 80860 Istanbul, Turkey} 

\date{\today} 
\draft

\twocolumn[\hsize\textwidth\columnwidth\hsize\csname@twocolumnfalse\endcsname
\maketitle 

\begin{abstract}
The effects of dilute Zn impurities on the uniform magnetic
susceptibility are calculated in the normal metallic state 
for a model of the spin fluctuations of the layered cuprates.
It is shown that scatterings from extended 
impurity potentials can 
lead to a coupling of the ${\bf q}\sim(\pi,\pi)$ and the 
${\bf q}\sim 0$ components of the magnetic susceptibility
$\chi({\bf q})$.
Within the presence of antiferromagnetic correlations, 
this coupling can enhance the uniform susceptibility.
The implications of this result for the experimental data 
on Zn substituted YBa$_2$Cu$_3$O$_{7-\delta}$ are discussed.
\end{abstract}

]

Experiments on Zn substituted cuprates 
provide valuable information 
on the magnetic properties of the host materials.
Measurements have shown that 
Zn impurities cause an enhancement of the uniform magnetic 
susceptibility \cite{Mahajan,Mendels}.
From the nearly perfect Curie-Weiss temperature dependence of the 
enhancement, the magnitude of the effective moment,
$\mu_{\rm eff}$, forming around the impurities 
has been extracted.
In underdoped YBa$_2$Cu$_3$O$_{6.66}$, 
$\mu_{\rm eff}$ is close to $\mu_{\rm B}$,
while in YBa$_2$Cu$_3$O$_7$
it is smaller by about a factor of 2.5.
Neutron scattering experiments have also shown that Zn 
impurities modify the spectrum of the magnetic fluctuations 
in these systems \cite{Sidis,Fong}.
Numerical calculations have emphasized the importance of the
antiferromagnetic correlations of the host in determining the 
effects of Zn substitution \cite{Poilblanc}. 
The effects of Zn impurities
have been also considered in various 
spin-gapped and spin-gapless antiferromagnetic models \cite{Sandvik}.
Furthermore,
studies have been carried out for the 
spin-gapped underdoped phase 
of the cuprates \cite{Gabay,Khaliullin}
as well as for the $d$-wave superconducting phase 
\cite{Quinlan,Li}. 
Here results will be presented for the normal state of the 
layered cuprates where there are short-range
antiferromagnetic correlations. 
Specifically, the effects of the
impurity scattering on the magnetic susceptibility 
will be calculated first
for a single impurity and then it will be scaled 
to an impurity concentration of 0.5$\%$
in the dilute limit.
It will be shown that 
scatterings from an extended impurity potential can 
lead to a coupling of the 
${\bf q}\sim 0$ and the ${\bf q}\sim(\pi,\pi)$ 
components of $\chi({\bf q})$.
Because of this coupling, the uniform susceptibility 
can get enhanced as antiferromagnetic correlations 
develop in the system.
In order to have a better understanding
of this process,
results will be presented on 
the related problem of 
how the ${\bf q}\sim 0$ and the 
${\bf q}\sim (\pi,\pi)$ components of $\chi({\bf q})$
get coupled by
a staggered charge-density-wave (CDW) field.

The starting point is the two-dimensional Hubbard model 
with an additional term 
representing the interaction of the electrons with a single 
impurity located at site ${\bf r}_0$,
\begin{eqnarray}
\label{Hubbard}
H=-t\sum_{\langle i,j\rangle ,\sigma} 
(c^{\dagger}_{i\sigma}c_{j\sigma}
+c^{\dagger}_{j\sigma}c_{i\sigma})
+ U \sum_i 
c^{\dagger}_{i\uparrow} c_{i\uparrow} 
c^{\dagger}_{i\downarrow} c_{i\downarrow}  \nonumber \\
-\mu \sum_{i,\sigma}
c^{\dagger}_{i\sigma} c_{i\sigma}
+
\sum_{i,\sigma} V_{\rm eff}({\bf r}_0,{\bf r}_i) 
c^{\dagger}_{i\sigma} c_{i\sigma}. 
\end{eqnarray}
Here $c_{i\sigma}$ ($c^{\dagger}_{i\sigma}$)
annihilates (creates) an electron with spin $\sigma$
at site ${\bf r}_i$,
$t$ is the near-neighbor hopping matrix element,
$U$ is the onsite Coulomb repulsion, and $\mu$ is the 
chemical potential.
The effective impurity-electron scattering potential is given by
$V_{\rm eff}({\bf r}_0,{\bf r}) =\sum_{\nu} V_{\nu} 
\sum_{{\bf \rho}_{\nu}} 
\delta({\bf r},{\bf r}_0+{\bf \rho}_{\nu})$,
where 
${\bf \rho}_{\nu}$ denotes the sites at a distance 
$\nu$ from the impurity site ${\bf r}_0$.
Here, the effective interaction is assumed to be  
static with a finite range extending 
a few lattice spacings away from the impurity.
The importance of using an extended impurity potential 
has been previously noted \cite{Xiang,Ziegler}.
For simplicity, in the following the hopping $t$ and 
the lattice constant $a$ will be set to 1.

\begin{figure} [b]
\begin{center}
\leavevmode
\epsfxsize=7.5cm 
\epsfysize=5cm 
\epsffile{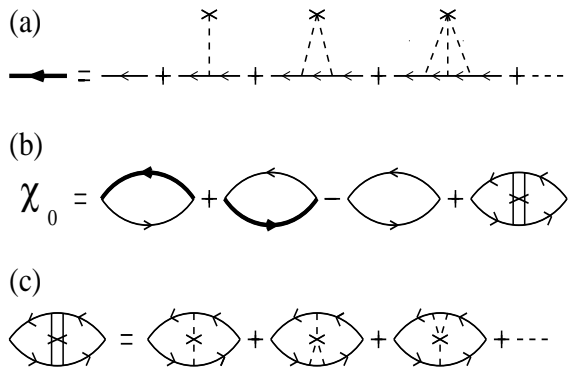}
\end{center}
\caption{
Feynman diagrams for (a) the dressed single-particle 
Green's function $G$, and 
(b)-(c) the irreducible susceptibility $\chi_0$
within the presence of a single impurity.
}
\label{fig1}
\end{figure}

The single-particle Green's function is given by
\begin{equation}
G({\bf r}_i,{\bf r}_j,i\omega_n) = 
- \int_0^{\beta} d\tau \, 
e^{i\omega_n \tau}
\langle c_{i\sigma}(\tau) 
c^{\dagger}_{j\sigma}(0) \rangle,
\label{G}
\end{equation}
where $\omega_n=(2n+1)\pi T$.
In the pure system with $U=0$,
one has in wavevector space
$G_0({\bf p},i\omega_n)=(i\omega_n-\varepsilon_{\bf p})^{-1}$
where
$\varepsilon_{\bf p} = -2t (\cos{p_x} + \cos{p_y}) - \mu$.
If a single impurity is introduced at site 
${\bf r}_0$, then 
\begin{eqnarray}
\label{GT}
G && ({\bf r},{\bf r'},i\omega_n) = 
G_0({\bf r},{\bf r'},i\omega_n)  \nonumber \\
&& + 
\sum_{{\bf r}'',{\bf r}'''} 
G_0({\bf r},{\bf r}'',i\omega_n) 
T({\bf r}'',{\bf r}''',i\omega_n)
G({\bf r}''',{\bf r}',i\omega_n), 
\end{eqnarray}
where the $T$-matrix for the impurity scattering is 
\begin{eqnarray}
\label{Tmatrix}
T && ({\bf r},{\bf r'},i\omega_n)=
\delta({\bf r},{\bf r'}) 
V_{\rm eff}({\bf r}_0,{\bf r}) \nonumber \\
&& + 
\sum_{{\bf r''}} V_{\rm eff}({\bf r}_0,{\bf r})
G_0({\bf r},{\bf r''},i\omega_n) 
T({\bf r''},{\bf r'},i\omega_n). 
\end{eqnarray} 
This is illustrated diagrammatically in Fig.~\ref{fig1}(a).
The magnetic susceptibility is defined as
\begin{equation}
\chi({\bf r},{\bf r'},i\omega_m) = 
\int_0^{\beta} d\tau \, 
e^{i\omega_m \tau}
\langle 
m^{-}({\bf r},\tau)  m^{+}({\bf r'},0) 
\rangle,
\label{chi}
\end{equation}
where $m^{+}(\bf r)=c^{\dagger}_{{\bf r}\uparrow} 
c_{{\bf r}\downarrow}$,
$m^{-}({\bf r})=c^{\dagger}_{{\bf r}\downarrow} 
c_{{\bf r}\uparrow}$, 
and $\omega_m=2m\pi T$. 
First, the effects of a single impurity 
will be calculated for $U=0$,
giving the irreducible susceptibility $\chi_0$, 
and 
then the effects of the Coulomb correlations will 
be included.
The diagrams representing the effects of a single impurity 
are shown in Figs.~\ref{fig1}(b) and (c).
Both the self-energy and the vertex corrections 
need to be included \cite{Langer}, and the 
resulting expression for 
$\chi_0({\bf r},{\bf r}')=
\chi_0({\bf r},{\bf r}',i\omega_m=0)$ 
is 
\begin{eqnarray}
\label{chi0}
\chi_0({\bf r},{\bf r'}) = - T \sum_{i\omega_n} 
\bigg\{
\big[ G({\bf r},{\bf r}', i\omega_{n}) 
G_0({\bf r}',{\bf r},i\omega_n) \nonumber \\
+
G_0({\bf r},{\bf r}',i\omega_{n}) 
G({\bf r}',{\bf r},i\omega_n)
-
(G_0({\bf r},{\bf r}',i\omega_{n}))^2 \big] \nonumber \\
+
\sum_{{\bf r}_1,{\bf r}_2,{\bf r}_3,{\bf r}_4}
G_0({\bf r},{\bf r}_1, i\omega_{n}) 
G_0({\bf r}_2,{\bf r}',i\omega_{n}) 
T({\bf r}_1,{\bf r}_2,i\omega_{n}) \nonumber \\
\times 
T({\bf r}_3,{\bf r}_4,i\omega_n)
G_0({\bf r}_3,{\bf r},i\omega_n) 
G_0({\bf r}',{\bf r}_4,i\omega_n) \bigg\}. 
\end{eqnarray}
Here the self-energy corrections are included 
by summing over the terms in the square brackets
rather than simply summing over $GG$.
In addition, the irreducible impurity-scattering 
vertex has been used 
in calculating the vertex corrections to $\chi_0$
instead of the reducible one.
These are necessary in order to prevent double counting when 
calculating the effects of just one impurity on 
$\chi_0$.

The Coulomb correlations are included 
by using the random-phase 
approximation, 
\begin{equation} 
\chi({\bf r},{\bf r}') = 
\chi_0({\bf r},{\bf r}') 
+ 
U \sum_{{\bf r}''}
\chi_0({\bf r},{\bf r}'')
\chi({\bf r}'',{\bf r}'). 
\label{rpa}
\end{equation}
Upon solving for $\chi({\bf r},{\bf r}')$,
the Fourier transform is taken, leading to 
$\chi({\bf q},{\bf q}')$.
At this point the impurity averaging can be done.
This restores the translational invariance, 
and the diagonal susceptibility 
$\chi({\bf q})=
\delta_{{\bf q}{\bf q}'} \chi({\bf q},{\bf q}')$
is obtained.
In order to minimize the finite size 
effects on $\chi({\bf q})$,
the single-particle Green's 
functions used in Eqs.~(\ref{GT}) and (\ref{Tmatrix}) 
are evaluated on large lattices basically 
without any finite size effects.
Equation~(\ref{rpa}), on the 
other hand, is solved on smaller 
$L\times L$ lattices for a single 
impurity located at the center,
but the resulting $\chi({\bf q})$ is
\noindent 
scaled to the $L\rightarrow \infty$ limit
while keeping the impurity concentration, 
$n_{\rm imp}$, fixed.
This is an efficient way of 
controlling the finite size effects
when working in the dilute limit. 
The following results on 
$\chi({\bf q})$
are for $n_{\rm imp}=0.005$, $U=2t$, 
electron filling 
$\langle n\rangle=0.86$, and $T=0.1$.
For these values of $U$ and $\langle n\rangle$,
the pure system has short-range 
antiferromagnetic fluctuations.
In addition,
the impurity potential was assumed to have a range 
of 2 lattice spacings with the 
following parameters:
$V_0=-20$, $V_1=0.5$, $V_{\sqrt{2}}=-0.5$,
and $V_2=-0.25$ \cite{Veff}.

\begin{figure} [b]
\begin{center}
\leavevmode
\epsfxsize=9cm 
\epsfysize=10cm 
\epsffile{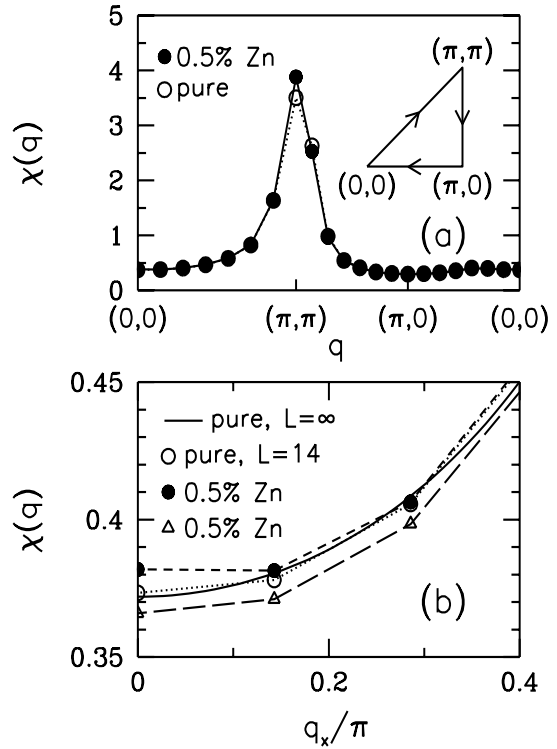}
\end{center}
\caption{
(a) $\chi({\bf q})$ versus ${\bf q}$ along the path  
shown in the inset for the pure and the 
0.5\% Zn substituted systems
on the $14\times 14$ lattice.
(b) $\chi({\bf q})$ versus $q_x$ along $q_x=q_y$.
The solid line is for the pure case on an infinite 
lattice.
The open and the filled circles are for 
the pure and the impure cases, respectively, 
on the $14\times 14$ lattice.
One observes that $\chi({\bf q}\rightarrow 0)$ is enhanced 
by the impurity scattering.
The triangles are also for the impure case
on the $14\times 14$ lattice, 
but now the impurity averaging has been done before 
including the Coulomb correlations.
This neglects the mixing of the ${\bf q}\sim 0$ and the 
${\bf q}\sim (\pi,\pi)$ components of $\chi({\bf q})$,
and subsequently $\chi({\bf q})$ is suppressed.
}
\label{fig2}
\end{figure}

Figure~\ref{fig2}(a) compares $\chi({\bf q})$ versus ${\bf q}$ 
for the pure and the 0.5$\%$ Zn substituted cases
obtained by using a 
$14 \times 14$ lattice in Eq.~(\ref{rpa}).
Figure \ref{fig2}(b) shows results on $\chi({\bf q})$ 
versus $q_x$ along $q_x=q_y$. 
Here the solid line represents $\chi({\bf q})$ 
for an infinite pure lattice, 
and the open circles are 
for the $14\times 14$ pure lattice. 
One sees that the finite size effects are small. 
The filled circles are the results obtained 
on the $14\times 14$ lattice for $n_{\rm imp}=0.005$. 
Comparing with the open circles, 
one observes that $\chi({\bf q}\rightarrow 0)$ is enhanced 
\noindent 
by the impurity scattering.
This can be understood better by considering
Eq.~(\ref{rpa}) in wavevector space,
\begin{equation} 
\chi({\bf q},{\bf q}') = 
\chi_0({\bf q},{\bf q}') 
+ 
U \sum_{{\bf q}''}
\chi_0({\bf q},{\bf q}'')
\chi({\bf q}'',{\bf q}').
\label{rpaq}
\end{equation}
Here,
one sees that once the translational invariance is lost, 
the Coulomb interaction mixes the different wavevector components of 
the magnetic susceptibility. 
This allows for the ${\bf q}\sim 0$
component to be influenced by 
the antiferromagnetic correlations. 
Through this mixing, 
$\chi({\bf Q}=(\pi,\pi))$ also gets enhanced 
as seen in Fig.~\ref{fig2}(a). 
The open triangles in Fig.~\ref{fig2}(b) 
are for the impure case, too,
but here the impurity averaging has been done 
before including the Coulomb correlations.
This does not allow for 
the mixing of the ${\bf q}\sim 0$  and the 
${\bf q}\sim (\pi,\pi)$ components, and consequently 
$\chi({\bf q})$ 
is suppressed.

\begin{figure} [b]
\begin{center}
\leavevmode
\epsfxsize=9.8cm 
\epsfysize=11.5cm 
\epsffile{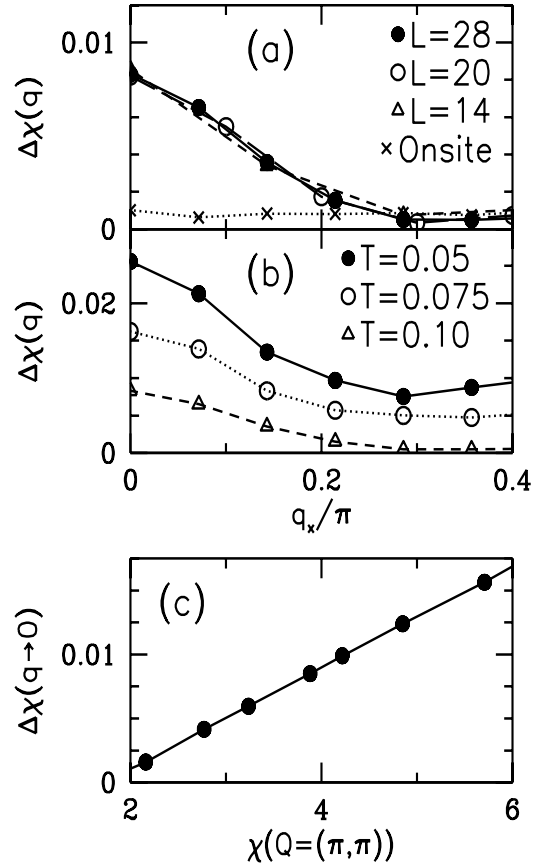}
\end{center}
\caption{
(a) Finite size effects on $\Delta\chi({\bf q})$ versus $q_x$
along $q_x=q_y$ for the extended impurity potential.
For comparison, $\Delta\chi({\bf q})$ obtained using an onsite 
impurity potential is shown by crosses, in which case 
$\Delta\chi({\bf q}\rightarrow 0)$ is smaller.
(b) Temperature evolution of $\Delta\chi({\bf q})$
versus $q_x$.
(c) $\Delta\chi({\bf q}\rightarrow 0)$ versus
$\chi({\bf Q}=(\pi,\pi))$ obtained by varying $U$ 
by small amounts,
which shows that these two quantities are correlated.
}
\label{fig3}
\end{figure}

Taking the difference of the filled and the empty circles 
shown in Fig.~\ref{fig2}(b), one obtains the enhancement of 
the magnetic 
susceptibility by the impurity scattering, 
$\Delta\chi({\bf q})$.
In Fig.~\ref{fig3}(a), 
$\Delta\chi({\bf q})$ versus $q_x$ is shown as the system
size is scaled from $14\times 14$ to 
$28\times 28$
while keeping $n_{\rm imp}$ fixed at 0.005.
One observes that the finite size effects on 
$\Delta\chi({\bf q})$ are negligible.
The inverse of the 
peak width of $\Delta\chi({\bf q})$ 
gives the size of the region around 
the impurity which contributes 
to the enhancement of the uniform susceptibility.
For comparison, 
$\Delta\chi({\bf q})$ obtained using 
an onsite impurity potential of
zero range with $V_0=-20$ 
is shown by crosses in Fig.~\ref{fig3}(a).
In this case,
$\Delta\chi({\bf q})$ is significantly smaller 
and it is featureless in ${\bf q}$.
Figure~\ref{fig3}(b) shows 
the temperature evolution of 
$\Delta\chi({\bf q})$
for the extended potential.
Here the growth of 
$\Delta\chi({\bf q}\rightarrow 0)$ with decreasing $T$ 
is seen.
On the other hand, 
for the onsite potential 
$\Delta\chi({\bf q})$ has a weak $T$ dependence
(not shown here).

In order to display the coupling of the ${\bf q}\sim 0$ and the 
${\bf q}\sim (\pi,\pi)$ components of $\chi({\bf q})$
by impurity scattering, 
the Coulomb repulsion $U$ entering Eq.~(\ref{rpa})
has been varied by small amounts while keeping 
the rest of the parameters fixed.
The resulting $\Delta\chi({\bf q}\rightarrow 0)$ versus 
$\chi({\bf Q}=(\pi,\pi))$ 
is plotted in Fig.~\ref{fig3}(c), where a linear 
dependence between these two quantities 
is seen. 
For instance,
by increasing $U$ by $4\%$ from 2.0 to 2.08, 
$\chi({\bf Q}=(\pi,\pi))$ increases from 3.9 to 5.7, 
while $\Delta\chi({\bf q}\rightarrow 0)$ doubles.
These calculations were repeated using 
various other sets of $V_{\nu}$'s 
and similar results were obtained
for the coupling between the ${\bf q}\sim 0$ and 
the ${\bf q}\sim (\pi,\pi)$ components of $\chi({\bf q})$.

The enhancement of $\chi({\bf q}\rightarrow 0)$ 
seen in Fig.~\ref{fig3}(a) is
about $2\%$.
However, a direct comparison with the experimental data
was not carried out
because of the simplicity of the model. 
For instance, the exact values of $V_{\nu}$'s are not known,
and 
the effects of the Coulomb correlations 
on the single-particle Green's functions and the $T$ dependence 
of $V_{\rm eff}$ are not taken into account.
For these reasons, 
it is not possible to make a direct comparison
with the Curie-Weiss behavior, either.

\begin{figure} [b]
\begin{center}
\leavevmode
\epsfxsize=6.5cm 
\epsfysize=4.2cm 
\epsffile{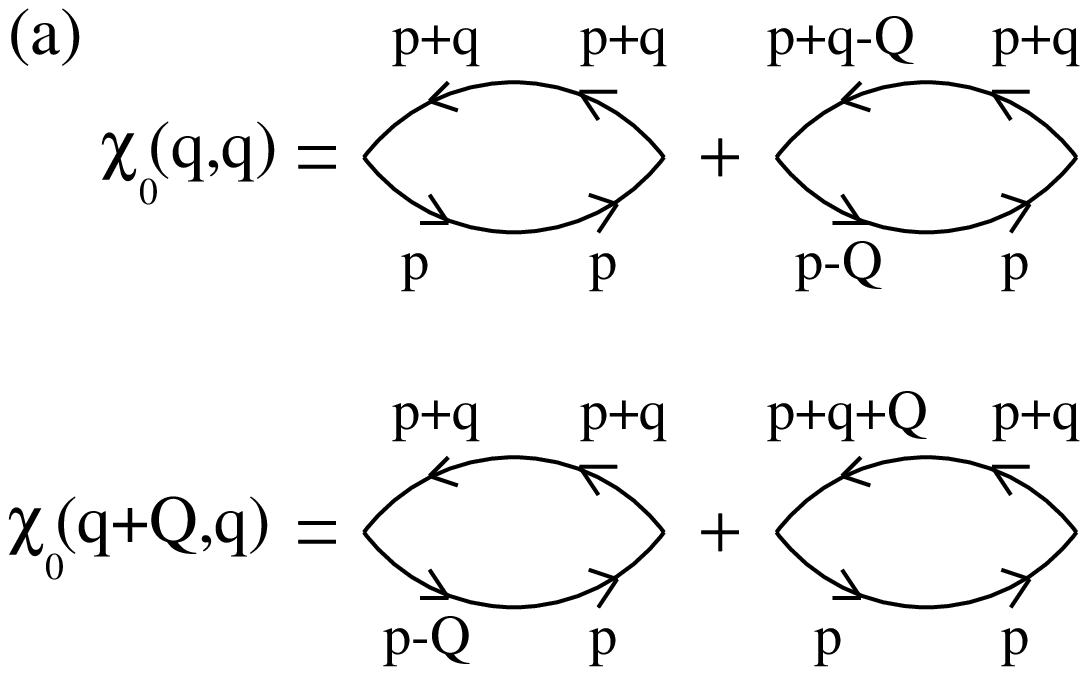}
\end{center}
\begin{center}
\leavevmode
\epsfxsize=9.6cm 
\epsfysize=5.6cm 
\epsffile[18 205 592 498] {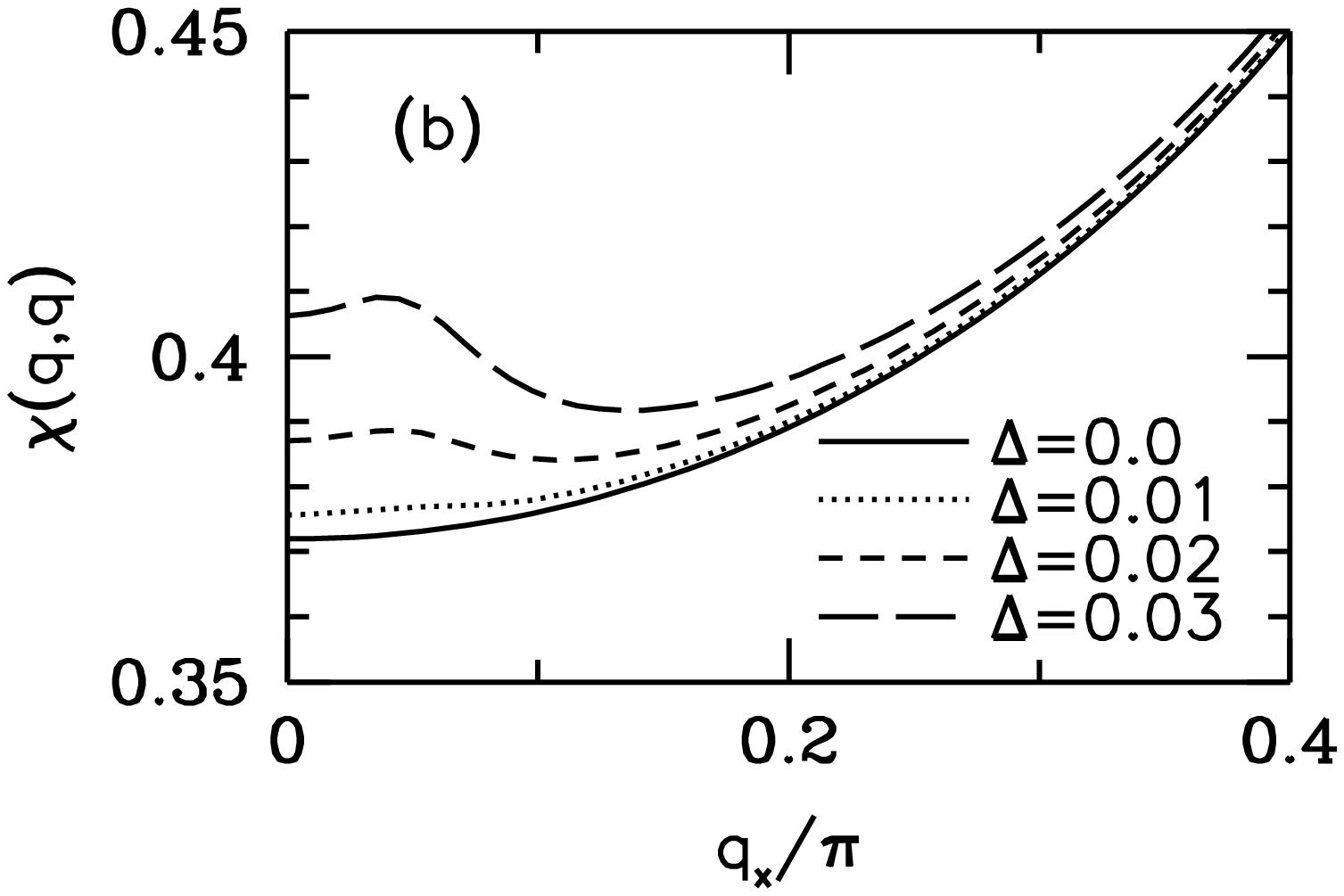}
\end{center}
\caption{
(a) Feynman diagrams representing the diagonal and the off-diagonal 
irreducible susceptibilities
$\chi_0({\bf q},{\bf q})$ and 
$\chi_0({\bf q}+{\bf Q},{\bf q})$,
respectively,
within the presence of a staggered CDW field.
(b) $\chi({\bf q},{\bf q})$ versus $q_x$ along $q_x=q_y$ for 
various values of the CDW amplitude $\Delta$,
where the enhancement of 
$\chi$ by $\Delta$ is seen.
}
\label{fig4}
\end{figure}

These results on the effects of scattering 
from extended 
impurity potentials
can be understood better if one considers 
how a staggered CDW field 
affects $\chi({\bf q})$
in a pure system.
This is a simple example 
which involves the mixing of the 
${\bf q}$ and the ${\bf q}+(\pi,\pi)$
components of $\chi({\bf q})$.
In fact, the calculation of $\chi({\bf q})$ for a single impurity 
reduces to this problem, if one only keeps 
the component of the effective impurity-electron 
interaction which transfers ${\bf Q}=(\pi,\pi)$ to the quasiparticles.
Here, one begins by introducing 
a staggered field $\Delta$ coupling 
to the site occupation number,
$\sum_{i\sigma} \Delta c^{\dagger}_{i\sigma} c_{i\sigma}
\exp{(i{\bf Q}\cdot{\bf r}_i)}$
with ${\bf Q}=(\pi,\pi)$.
The resulting irreducible susceptibility has both diagonal 
and off-diagonal terms, 
$\chi_0({\bf q},{\bf q})$ and $\chi_0({\bf q}+{\bf Q},{\bf q})$, 
which are shown diagrammatically in Fig.~\ref{fig4}(a). 
Within the random-phase approximation,
$\chi_0({\bf q},{\bf q})$ and 
$\chi_0({\bf q}+{\bf Q},{\bf q}+{\bf Q})$ 
are coupled through the 
off-diagonal term $\chi_0({\bf q}+{\bf Q},{\bf q})$
leading for $\chi({\bf q},{\bf q})$ to
\begin{eqnarray}
\label{rpacdw}
\chi({\bf q},{\bf q}) = 
D^{-1}
\big\{ \chi_0({\bf q},{\bf q})(1-U\chi_0({\bf q}
+{\bf Q},{\bf q}+{\bf Q}))  \nonumber \\
+ U\chi_0^2({\bf q}+{\bf Q},{\bf q})  \big\}, 
\end{eqnarray}
where
$D=(1-U\chi_0({\bf q},{\bf q}))(1-U\chi_0({\bf q}
+{\bf Q},{\bf q}+{\bf Q}))
- U^2\chi_0^2({\bf q}+{\bf Q},{\bf q})$.
This calculation of $\chi$ within a CDW field 
is similar to that within
a spin-density-wave field \cite{Schrieffer}.
The irreducible susceptibilities seen in Fig.~\ref{fig4}(a) can be 
evaluated and the resulting $\chi({\bf q},{\bf q})$ 
is displayed in Fig.~\ref{fig4}(b), where the enhancement of 
$\chi({\bf q},{\bf q})$ by $\Delta$ is seen. 
It can be shown that, for small $\Delta$, 
the off-diagonal term $\chi_0({\bf q}+{\bf Q},{\bf q})$ 
remains finite in the limit 
${\bf q}\rightarrow 0$ \cite{CDW},
and it is approximately equal to 
$(\Delta/\mu)\chi_0({\bf q},{\bf q})$, while 
$\chi_0({\bf q},{\bf q})$ is changed little by $\Delta$.
Hence, it is expected from Eq.~(\ref{rpacdw}) 
that the uniform susceptibility 
will get enhanced by the antiferromagnetic correlations,
as $\Delta$ is turned on. 
In addition, 
it can be shown that 
$\chi_0({\bf q}+{\bf Q},{\bf q})$ is 
a smooth function of ${\bf q}$ 
for ${\bf q}\sim 0$.
Hence, from Eq.~(\ref{rpacdw}) one observes that 
the ${\bf q}$ structure of the enhancement in 
$\chi({\bf q},{\bf q})$ for ${\bf q}\sim 0$ reflects 
the structure of
$\chi({\bf q},{\bf q})$ for ${\bf q}\sim (\pi,\pi)$,
which is slightly incommensurate 
at this temperature.

It is important to note that here 
the enhancement of the uniform susceptibility 
is not due to an enhancement of 
$\chi_0({\bf q},{\bf q})$ by $\Delta$, since these small
values of $\Delta$ have little effect on 
$\chi_0({\bf q},{\bf q})$. 
Rather, it is due to the nonvanishing of the anomalous 
susceptibility $\chi_0({\bf q}+{\bf Q},{\bf q})$, 
which allows for a coupling to the 
antiferromagnetic correlations. 
In 
\noindent the impurity problem, 
the enhancement of $\chi({\bf q}\sim 0)$ 
occurs similarly.
In Eq.~(\ref{rpaq}),
the off-diagonal terms $\chi_0({\bf q},{\bf q}''\neq{\bf q})$,
which are nonvanishing since the impurity scattering 
does not conserve momentum, 
couple the ${\bf q}\sim 0$ and the ${\bf q}''\sim (\pi,\pi)$ 
components of $\chi$, 
enhancing the uniform susceptibility.
This coupling is strong especially when 
an extended impurity potential is used.
However,
if the impurity averaging is done before 
including the Coulomb correlations, this coupling is neglected and 
$\chi({\bf q})$ is suppressed with respect to the pure case,
as it was seen in Fig. \ref{fig2}(b).

In summary, the effects of dilute Zn impurities on the 
uniform magnetic susceptibility have been 
calculated in 
a metallic model 
which has short-range antiferromagnetic correlations, 
but does not necessarily have a spin gap. 
It has been found that impurity scattering 
through an extended potential leads
to the mixing of the ${\bf q}\sim(\pi,\pi)$ and the ${\bf q}\sim 0$ 
components of $\chi({\bf q})$.
Because of this coupling, the antiferromagnetic correlations can
enhance the uniform susceptibility. 
The microscopic model presented here predicts 
a strong dependence of $\Delta \chi({\bf q}\rightarrow 0)$ on the 
strength of the antiferromagnetic correlations. 
This could play a role in determining the hole-doping 
and the temperature dependence of 
$\Delta \chi({\bf q}\rightarrow 0)$ 
in Zn substituted YBa$_2$Cu$_3$O$_{7-\delta}$.
However, it must be kept in mind that these results 
depend on the nature of the 
effective impurity interaction and on the 
validity of the approach used to treat the Coulomb correlations.


The author thanks H.F. Fong for helpful discussions.
The numerical computations reported in this paper were performed 
at the Center for Information Technology at Ko\c{c} University.



\begin{thebibliography}{999}

\bibitem{Mahajan} A.V. Mahajan {\it et al.},
\prl {\bf 72}, 3100 (1994).

\bibitem{Mendels} P. Mendels {\it et al.},
Europhys. Lett. {\bf 46}, 678 (1999).

\bibitem{Sidis} Y. Sidis {\it et al.},
\prb {\bf 53}, 6811 (1996).

\bibitem{Fong} H.F. Fong {\it et al.},
\prl {\bf 82}, 1939 (1999).

\bibitem{Poilblanc} D. Poilblanc {\it et al.},
\prl {\bf 72}, 884 (1994); 
\prb {\bf 50}, 13020 (1994).

\bibitem{Sandvik} A. Sandvik {\it et al.},
\prb {\bf 56}, 11701 (1997).

\bibitem{Gabay} M. Gabay,
Physica C {\bf 235}-{\bf 240}, 1337 (1994).

\bibitem{Khaliullin} G. Khaliullin {\it et al.},
\prb {\bf 56}, 11882 (1997).
 
\bibitem{Quinlan} S.M. Quinlan {\it et al.},
\prb {\bf 51}, 497 (1995).

\bibitem{Li} J.-X. Li {\it et al.},
\prb {\bf 58}, 2895 (1998).

\bibitem{Xiang} T. Xiang {\it et al.},
\prb {\bf 51}, 11721 (1995).

\bibitem{Ziegler} W. Ziegler {\it et al.},
\prb {\bf 53}, 8704 (1996).

\bibitem{Langer} J.S. Langer, 
Phys. Rev. {\bf 120}, 714 (1960).

\bibitem{Veff} These values for $V_{\nu}$ are comparable 
to those obtained in Ref.~\cite{Ziegler}.
The results presented here will not 
depend sensitively on the specific values of $V_{\nu}$ but 
on whether $V_{\rm eff}$ is extended or not.

\bibitem{Schrieffer} J.R. Schrieffer {\it et al.}, 
\prb {\bf 39}, 11663 (1989).

\bibitem{CDW} This is a consequence of the CDW coherence factors.

\end{thebibliography}
\end{document}